\input amssym.def
\input amssym.tex
\def\gr{$\gamma$-ray }
\def\grs{$\gamma$-rays }

\def\be{\begin{equation}}
\def\ee{\end{equation}}
\def\bdm{\begin{displaymath} }
\def\edm{\end{displaymath} }
\documentclass{aa}
\usepackage{times}
\usepackage{graphicx}
\begin{document}

\title{No evidence yet for hadronic TeV gamma-ray emission from SNR
RX~J1713.7$-$3946}
\author{O. Reimer\inst{1} \and M. Pohl\inst{1}}
\date{Received ? / Accepted ?}
\institute{Institut f\"ur Theoretische Physik, Lehrstuhl IV: 
Weltraum- und Astrophysik, Ruhr-Uni\-ver\-si\-t\"at Bochum, 
D-44780 Bochum, Germany 
}

\offprints{O. Reimer, olr@tp4.ruhr-uni-bochum.de}
\date{received/accepted}
\authorrunning{Reimer et al.}
\titlerunning{No hadronic TeV \grs from SNR RX~J1713.7$-$3946}
\abstract{Recent TeV-scale \gr 
observations with the CANGAROO II telescope have led to the claim that
the
multi-band spectrum of RX~J1713.7$-$3946 cannot be explained as the
composite 
of a synchrotron and an inverse Compton component emitted by a
population of 
relativistic electrons. It was argued that the spectrum of the
high-energy 
emission is a good match to that predicted by pion decay, thus providing 
observational evidence that protons are accelerated in SNR to at least
TeV energies.\\
In this Letter we discuss the multi-band spectrum of RX~J1713.7$-$3946
under
the constraint that the GeV-scale emission observed from the closely
associated
EGRET source 3EG~J1714$-$3857 is either associated with the SNR or an
upper limit 
to the gamma-ray emission of the SNR. \\
We find that the pion-decay model
adopted by Enomoto et al. is in conflict with the existing GeV data. We have examined the possibility of a modified proton spectrum to explain the data,
and find that we cannot do so within any existing theoretical framework of
shock acceleration models.
}
\maketitle

\section{Evidence for particle acceleration in SNR}
Supernova remnants (SNR) are considered the most likely sources of
cosmic rays, either
as individual accelerators or by their collective effect in
superbubbles. Yet
observational evidence in favor of this scenario has been found only for
cosmic-ray 
electrons, not for the nucleons. 

Three shell-type SNR have been detected at TeV \gr energies so far, all
of which
show non-thermal X-ray emission, which presumably is synchrotron
radiation.
It is known that the synchrotron radiating electrons would
inverse-Compton scatter
the microwave background and the ambient far-infrared photon field
to TeV \gr energies with a flux depending mainly on the X-ray flux and
the magnetic
field strength within the remnant \cite{po96}, provided both are
measured at photon
energies corresponding to the same electron energy and escape from the
compression
region at the SNR shock is inefficient. The X-ray and TeV spectra should
then be similar,
which would permit one to discriminate between a hadronic and a leptonic
origin of the radiation.

Indeed, the observed TeV \gr spectrum of SN~1006 
(Tanimori et al. \cite{ta98,ta01}) is consistent with
synchrotron/inverse Compton models (\cite{md96,aa99,nai99}). A
significant
contribution of \grs from hadronic interactions appears unlikely on
account of
the low density environment in which the remnant resides.

Recently, TeV \grs have been detected from Cassiopeia A \cite{aha01},
if with 0.03 Crab above 1 TeV at a flux much lower that that reported
for SN~1006.
Estimates for the magnetic field strength
at the shock and in the downstream region of Cas~A suggest, that the
leptonic TeV-scale
\gr emission should be substantially weaker than in case of SN~1006 and
should display a cut-off near 1 TeV. 
Because of the moderate statistical significance of the detection, the
\gr
spectral index is poorly constrained. Thus the present data can be
interpreted 
as leptonic or hadronic \gr emission or as a mixture of both.

The SNR RX~J1713.7$-$3946 is in many respects similar to SN~1006.
Observations have 
revealed intense, apparently non-thermal X-ray emission from the
north-west rim
\cite{koy97}, and TeV \gr emission from that region \cite{mur00}. 
Recent measurements with the CANGAROO II telescope have indicated that
the
TeV-scale \gr spectrum of RX~J1713.7$-$3946 can be well represented by a
single
power-law with index $\alpha \simeq 2.8$ between 400~GeV and 8~TeV
\cite{eno02}.
The authors argue that the multi-band spectrum from
radio frequencies to TeV \gr energies cannot be explained as the
composite 
of a synchrotron and an inverse Compton component emitted by a
population of 
relativistic electrons. It is then claimed that the spectrum of the
high-energy 
emission is a good match to that predicted by pion decay. Hence
RX~J1713.7$-$3946 
would provide observational evidence that protons are accelerated in SNR
to at 
least TeV energies.

In this Letter we discuss the multi-band spectrum of RX~J1713.7$-$3946
under
the constraint that the GeV scale emission observed from the closely
located
EGRET source 3EG~J1714$-$3857 is taken into account as either being
associated 
with the SNR or an upper limit to the emission of the SNR. 
For both cases we find that a pion decay origin of the observed TeV-scale \gr
emission of RX~J1713.7$-$3946 is highly unlikely, and propose further
observations which should help arrive at a solution.

\section{The multi-band spectrum of RX~J1713.7$-$3946 revisited}

\begin{figure}
\resizebox{\hsize}{!}{\includegraphics{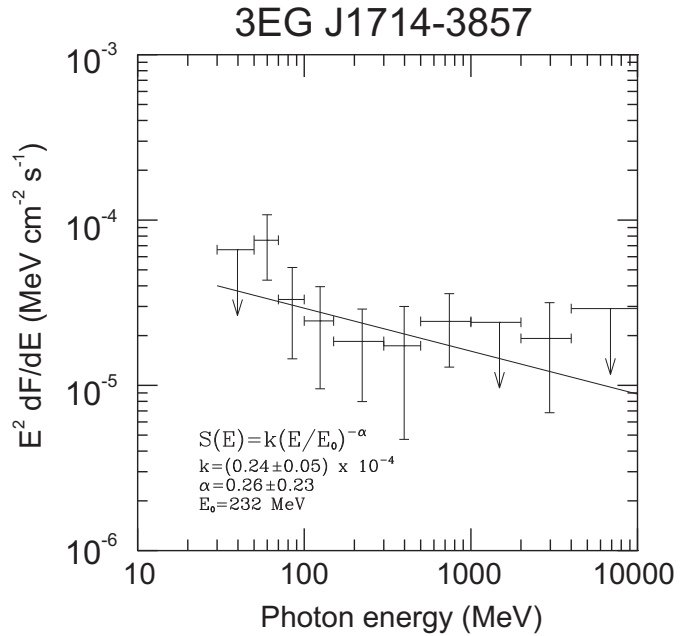}}
\caption{The \gr spectrum of 3EG~J1714$-$3857 between 10 MeV and 10 GeV as 
obtained for the 3EG catalog \cite{ha99}. The photon spectrum is
publicly 
accessible at the Compton Science Support Center at GSFC. The solid line
indicates
a power-law fit to the observed spectrum, whose parameters are given in
the figure.}
\label{spec}
\end{figure}

RX~J1713.7$-$3946 is located in the vicinity of a molecular cloud complex
(Slane et al. \cite{sla99}), 
which itself coincides with the unidentified GeV-scale
\gr source 3EG~J1714$-$3857 \cite{ha99}. The best-fit positions of
3EG~J1714$-$3857 and the CANGAROO TeV \gr source are about
$0.7\degr$ apart. The 2$\sigma$ positional uncertainties toward each
other are
$0.4\degr$ for the EGRET source and $> 1\degr$ for the CANGAROO source. 
Nevertheless, the two experiments may have
observed the same source, though that is not very likely on account of
the spatial
separation.  
In Fig.\ref{spec} we show the \gr spectrum of 3EG J1714$-$3857 as
determined during the
compilation of the Third EGRET catalog \cite{ha99} and since that time
publicly available 
at the Compton Science Support Center.

If RX~J1713.7$-$3946 and 3EG~J1714$-$3857 indeed represent the same source
seen at different 
wavelengths, the actual observed GeV-scale emission of 3EG~J1714$-$3857
must be reproduced by
any viable model for RX~J1713.7$-$3946. 
If the two are different sources, the GeV-scale \gr radiation of
RX~J1713.7$-$3946 must 
be less than the emission observed from the EGRET source. In any event,
the GeV-scale
\gr radiation emitted by RX~J1713.7$-$3946 cannot exceed that observed
from 3EG~J1714$-$3857.

\begin{figure*}
\centering
 \includegraphics[width=17cm]{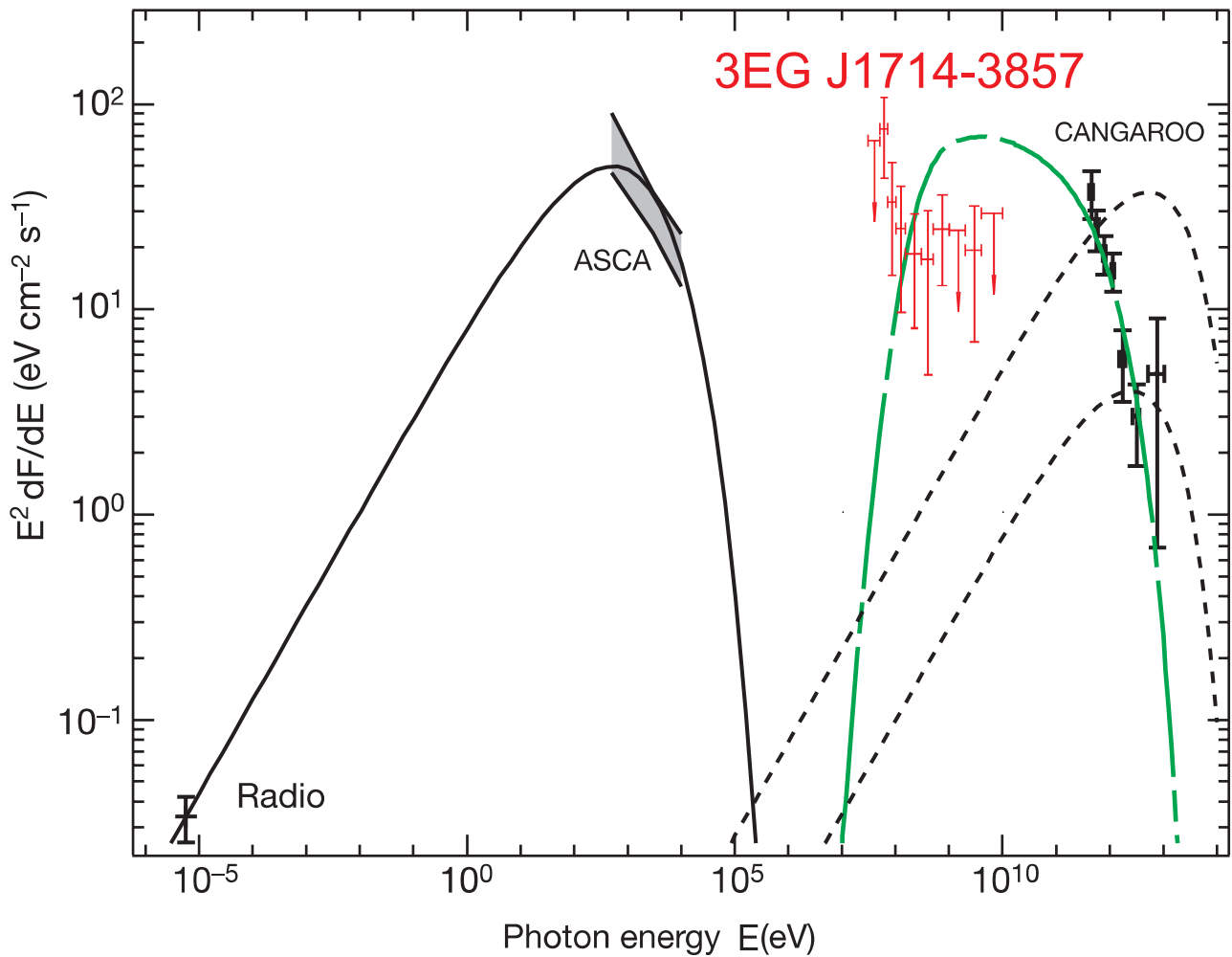}
 \caption{The multi-band spectrum of RX J1713.7$-$3946 from Enomoto et al.
(2002)
revised to include the \gr spectrum of 3EG~J1714$-$3857, shown in
comparison with emission
models presented in the same publication. The solid line indicates the
synchrotron 
emission of relativistic electrons, and the dashed lines are the
corresponding
inverse Compton spectra based on the microwave background and the
ambient 
far-infrared photon field for two sets of parameters. As Enomoto et al.
have
pointed out, both IC model spectra violate the observed TeV \gr
spectrum.
The green (grey) long-short-dashed line shows the pion decay spectrum,
which
was supposed to reproduce the \gr emission of RX~J1713.7$-$3946. The
modelled GeV-scale
pion decay flux would significantly exceed the observed emission by a
factor of three.}
 \label{mfs}
\end{figure*}

In Fig.\ref{mfs} we show the multi-band spectrum of RX~J1713.7$-$3946,
originally
presented by Enomoto et al. (2002), here modified to include the \gr
spectrum of 3EG~J1714$-$3857. To be noted from the figure is that the
predicted
GeV-scale flux from $\pi^0$-decay significantly exceeds the observed
flux,
thus prohibiting a hadronic origin of the TeV-scale \gr emission from
RX~J1713.7$-$3946.

\section{Discussion}

Is it possible to modify the proton spectrum such that the resultant
pion-decay \gr spectrum complies with the observed multiband spectrum
that is shown in Fig.\ref{mfs}?

Enomoto et al. have used a power-law proton spectrum with high-energy cut-off
($s=2.08, E_{\rm c}\simeq 50$~TeV) to calculate the \gr spectrum. 
We have used a scaling model for the nucleon-nucleon interactions
(B\"usching et al. \cite{bue01}) to realistically
investigate what proton spectra would have a \gr yield in accord
with that observed. As a result we find that a low-energy cut-off must 
be imposed on the proton spectrum with 
$E_1\gtrsim 100$~GeV (for s=2). We do not know a process that would cause
such a low-energy cut-off in the spectrum of particles accelerated by a SNR.

Alternatively, a broken power-law ($s_1 \lesssim 1.6$, $s_2 \simeq 2.7$, $E_2
\simeq 20$~TeV) would fit the data, whereas a hard power-law with
exponential cut-off gives a very bad fit.
We note that such a broken power-law spectrum is not predicted by 
shock acceleration models and is also in conflict with the spectra
observed from accelerated electrons in SNR. Additionally, the necessity to
introduce many parameters to explain a few data points hardly qualifies
as evidence for the underlying model, even more so if these parameters are
far away from what can be expected based on observations and theoretical 
modelling.
Therefore we conclude that
a pion decay origin of the observed TeV-scale \gr
emission of RX~J1713.7$-$3946 is highly unlikely. 

Having established that pion decay is not a viable model for the TeV \gr
emission
from RX~J1713.7$-$3946, contrary to the proposal by Enomoto et al. (2002), 
one has to reconsider the possible association of the EGRET and the
CANGAROO source as well as
the origin of the multi-band emission. 

The low intensity of the thermal component determined in thermal
emission
plus power-law spectral fits to X-ray data of SNR RX~J1713.7$-$3946
(Slane et al. \cite{sla99}, Pannuti \& Allen 
\cite{pa02}) suggests that the remnant is still expanding into the wind
bubble of the 
progenitor \cite{ell01}. No analysis of X-ray data of the possible
interaction region
between the molecular clouds and either the progenitor's stellar wind or
the
supernova blast wave has been published so far.
It has been proposed that the radio, X-ray, and TeV \gr emission of RX
J1713.7$-$3946
is produced by synchrotron radiation and inverse Compton scattering by
relativistic
electrons, similar to the case of SN~1006, but that the EGRET source
should be
identified with the nearby molecular cloud, which provides dense target
material for 
relativistic protons having escaped from the SNR (Slane et al.
\cite{sla99}, Butt et al. \cite{bu01}). 

While the spatial
association of the EGRET and the CANGAROO source can be observationally
tested 
with forthcoming \gr missions such as GLAST, the expected absolute flux 
level of $\pi^0$-decay \grs is not well determined in the context of
general 
acceleration models (for a detailed discussion see Pohl (\cite{po02})). 
One of the crucial but poorly known parameters is the 
injection efficiency, with which suprathermal protons are injected at 
the shock front. In contrast to the electrons, for which the non-thermal
X-ray flux can be used as a primer for the electron flux, whatever the
micro-physics at the acceleration site, the high energy cosmic ray 
nucleons do not reveal themselves in any presently observable channel
other than
\gr emission from $\pi^0$-decay and radiation from secondary electrons. 
The overprediction of the hadronic TeV \gr flux from 
Cas~A and the Tycho SNR (Atoyan et al. \cite{at00}, V\"olk et al.
\cite{voe01})
also compromises corresponding model
predictions for hadronic \gr emission from SN~1006 \cite{bere02}.

The question whether or not inverse Compton scattering can be
responsible for the
observed TeV-scale \gr emission of RX~J1713.7$-$3946, whatever the origin
of the EGRET
source 3EG~J1714$-$3857, will be easier to answer once a better defined
synchrotron spectrum is available. It is important to know at what
frequency the peak of the synchrotron intensity is located
in a $\nu F_\nu$ spectrum, for that determines the location
of the peak of the inverse Compton component. For the inverse Compton 
model to work, the peak of the synchrotron spectrum should be located at
about an order of magnitude lower frequency than in the best fit of
Enomoto et al., implying the power-law index $s\le 2.0$, which is not excluded 
by the sparse radio data available to date. The analysis
is complicated and clearly beyond the scope of this Letter, for the
radio image of SNR
RX~J1713.7$-$3946 displays small-scale structure with the high magnetic field
regions producing the bulk of the synchrotron radiation and the low
field regions presumably being responsible for most of the inverse Compton 
emission. Nevertheless, CHANDRA and NEWTON
observations in conjunction with radio measurements at different
frequencies
may be sufficient to prove or disprove the assertion that inverse
Compton models are not
viable for RX~J1713.7$-$3946.

\section{Summary}

In this Letter we have discussed the most recent TeV \gr measurements of
RX~J1713.7$-$3946
\cite{eno02} under
the constraint that the GeV scale emission observed from the closely
associated
EGRET source 3EG~J1714$-$3857 is either associated with the SNR or an
upper limit 
to the gamma-ray emission of the SNR.

Our conclusions are the following:

\begin{itemize}

\item 
the nearby EGRET source 3EG~J1714$-$3857 may or may not be related to 
RX~J1713.7$-$3946. In any event, the GeV-scale
\gr radiation emitted by RX~J1713.7$-$3946 cannot exceed that observed
from 3EG~J1714$-$3857.

\item 
a pion decay origin of the observed TeV-scale \gr
emission of RX~J1713.7$-$3946 is highly unlikely, contrary to a previous claim.

\item
answering the question whether or not inverse Compton scattering can be
responsible 
for the observed TeV-scale \gr emission of RX~J1713.7$-$3946, whatever the
origin of
the EGRET source 3EG~J1714$-$3857, requires a better knowledge of the
synchrotron spectrum
of the SNR than available to date.

\end{itemize}  

\begin{acknowledgements}
Financial support for OR and MP by the Bundesministerium f\"ur Bildung
und Forschung through 
DESY, grant {\it 05 CH1 PCA/6}, and DLR, grant {\it 50 QV 0002}, is
gratefully acknowledged.
We are indebted to Yousaf M. Butt, Diego F. Torres, and Gustavo 
E. Romero, who went along with this manuscript in frequent and extensive 
discussions. Due to an editorial request by Nature concerning overlapping 
authorlists between a brief communication to Nature by Butt et al. \cite{bu02}
and this letter they decided to obey and withdraw from coauthorship at a time 
when this letter has already been accepted for publication. 
\end{acknowledgements}

\end{document}